\documentclass[aps,pra,twocolumn,groupedaddress,showpacs]{revtex4}

\usepackage[dvips]{graphicx}
\usepackage{epsfig}
\usepackage{amssymb}
\usepackage{amsmath}
\usepackage{color}

\begin{document}


\title{Experimental demonstration of a Rydberg-atom beam splitter}


\author{J. Palmer and S. D. Hogan}

\affiliation{Department of Physics and Astronomy, University College London, Gower Street, London WC1E 6BT, U.K.}


\date{\today}

\begin{abstract} 
Inhomogeneous electric fields generated above  two-dimensional electrode structures have been used to transversely split beams of helium Rydberg atoms into pairs of spatially separated components. The atomic beams had initial longitudinal speeds of between 1700 and 2000~m/s and were prepared in Rydberg states with principle quantum number $n=52$ and electric dipole moments of up to 8700~D by resonance-enhanced two-color two-photon laser excitation from the metastable 1s2s\,$^3$S$_1$ level. Upon exiting the beam splitter the ensembles of Rydberg atoms were separated by up to 15.6~mm and were detected by pulsed electric field ionization. Effects of amplitude modulation of the electric fields of the beam splitter were shown to cause particle losses through transitions into unconfined Rydberg-Stark states. 
\end{abstract}


\maketitle

\section{INTRODUCTION}\label{sec:introduction}

A wide range of Rydberg atom and molecule optics elements, based on the interactions of samples in quantum states possessing large static electric dipole moments with inhomogeneous electric fields~\cite{wing80a,breeden81a}, have been developed in recent years. These include guides~\cite{lancuba13a,ko14a}, deflectors~\cite{townsend01a,allmendinger14a}, mirrors~\cite{vliegen06a}, lenses~\cite{vliegen06b}, decelerators~\cite{yamakita04a,vliegen04a,vliegen05a,hogan12a,lancuba14a} and traps~\cite{hogan08a,seiler11a,hogan12a,lancuba16a} composed of three-dimensional electrode structures, and chip-based electrode arrays. These devices have been employed, e.g., to study effects of blackbody radiation on the evolution of Rydberg-Stark states on time scales up to several milliseconds~\cite{seiler11a,seiler12a}, to characterize effects of collision-induced $m$-changing arising from dipole-dipole interactions in ensembles of polar Rydberg atoms~\cite{seiler16a}, to identify classes of long-lived molecular Rydberg states which are immune to fast predissociation~\cite{hogan09a,seiler11b}, to study ion-molecule reactions at low collision energies in merged beams~\cite{allmendinger16a,allmendinger16b}, and to exert control over the translational motion of neutral beams of exotic positronium atoms~\cite{deller16a}. 

Further experiments are under development~\cite{hogan16a} in which these Rydberg atom optics elements will be exploited for studies of molecular collisions and decay processes, for transport and confinement in hybrid approaches to quantum information processing~\cite{hogan12b}, and for investigations of the acceleration of particles composed of antimatter in the gravitational field of the Earth using positronium~\cite{cassidy14a} and antihydrogen~\cite{kellerbauer08a} atoms. In several of these areas methods for splitting beams of Rydberg atoms into spatially separated components promise to be of great value, (1) for the distribution of samples among spatially separated interaction regions,(2) for reference intensity measurements in collision experiments, and (3) if prepared at sufficiently low translational temperatures, for atom or molecule interferometry. 

With applications of this kind in mind we report here the experimental demonstration of a Rydberg-atom beam splitter, the operation of which relies upon the generation of carefully tailored inhomogeneous electric field distributions above two-dimensional arrays of metallic electrodes. This device complements recently developed chip-based electron beam splitters~\cite{hammer15a}, and beam splitters for polar ground state molecules composed of two-dimensional~\cite{deng11a} and three-dimensional electrode structures ~\cite{gordon16}.

\begin{figure}
\begin{center}
\includegraphics[width = 0.48\textwidth, angle = 0, clip=]{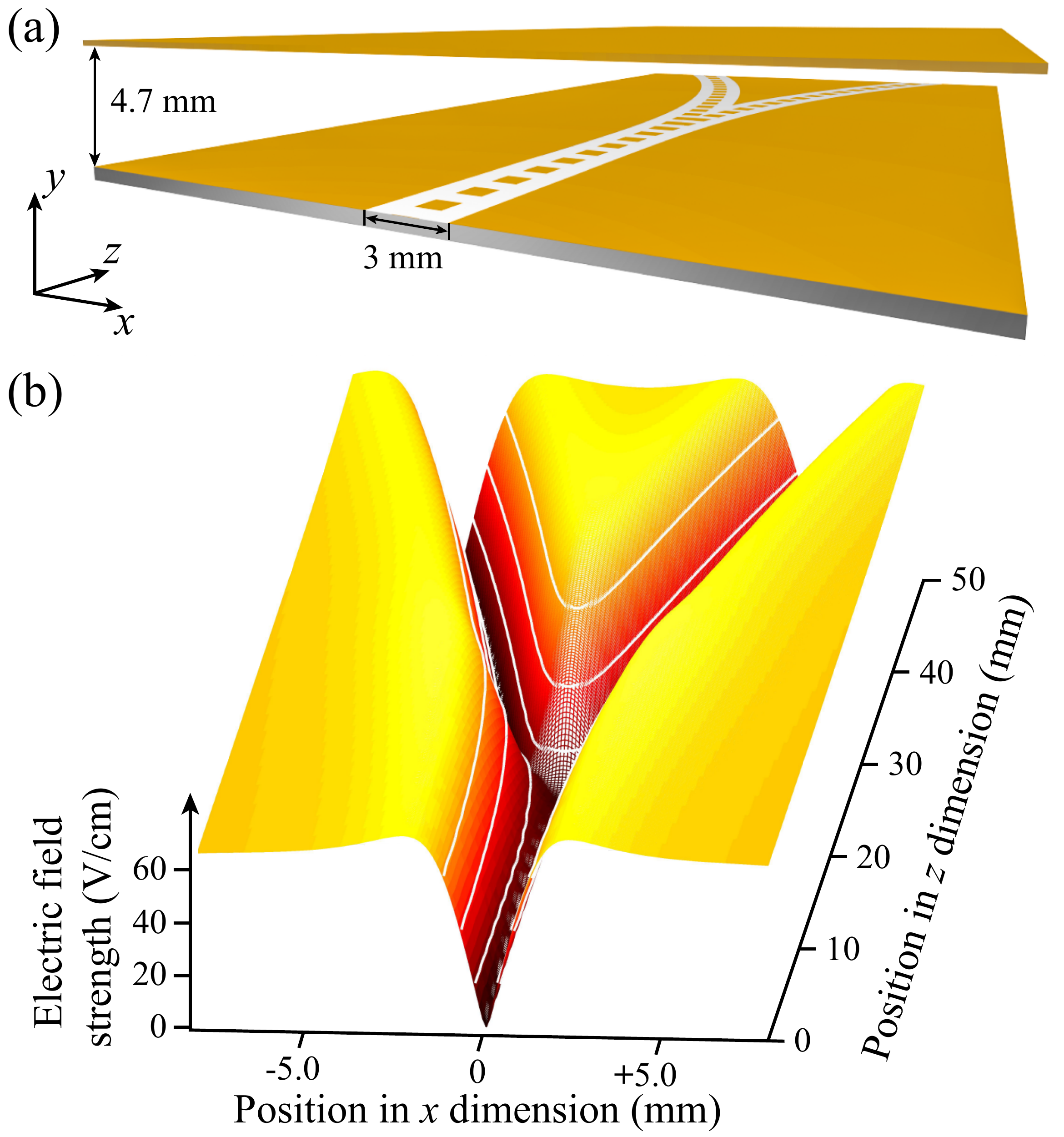}
\caption{(Color online) (a) Schematic diagram of a Rydberg atom beam splitter. (b) Electric field distribution in the $xz$ plane of the beam splitter at $y$ = 1.4~mm for  $V_{\mathrm{g}}=30$~V (see text for details). The electric field strength indicated by the white contour lines in (b) is displayed in intervals of 20~V/cm.}
\label{fig1}
\end{center}
\end{figure}

\section{BEAM SPLITTER DESIGN}\label{sec:design}

A schematic diagram of the Rydberg-atom beam splitter used in the experiments reported here is shown in Fig.~\ref{fig1}(a). The design of this device is based on the geometry of a coplanar microwave transmission line allowing it to be integrated with two-dimensional microwave circuits~\cite{hogan12b} and transmission-line guides~\cite{lancuba13a}, decelerators~\cite{lancuba14a}, and traps~\cite{lancuba16a} for beams of Rydberg atoms and molecules. The beam splitter consists of a 90-mm-long two-dimensional electrode array composed of a series of 1~mm $\times$ 1~mm square electrodes with a center-center separation of 2~mm, and a separation of 1~mm on each side from the adjacent ground planes. An upper plate electrode is positioned 4.7~mm above this array. The split section of the device is formed by the intersection of the arcs of two circles with radii of curvature of 0.3~m.

In the experiments the beam splitter was operated in an electrostatic mode. For a potential $V_{\mathrm{g}}$ applied to each electrode of the segmented center conductor of the transmission-line and the upper plate electrode, a two-dimensional quadrupole electric field distribution is generated in the $xy$ plane above the individual arms of the device [see Fig.~\ref{fig1}(a) for a definition of the coordinate system]. This quadrupole field acts as a guide for atoms in low-field-seeking Rydberg-Stark states with positive Stark energy shifts. As can be seen in Fig.~\ref{fig1}(a) the guide gradually splits into two spatially separated components with increasing longitudinal position above the electrode array, reflecting the change in structure of the center conductor electrodes. The calculated electric field distribution at a height of $y$ = 1.4~mm above the surface of the electrode array in the region where the paths diverge is displayed in Fig.~\ref{fig1}(b) for $V_{\mathrm{g}}$ = 30~V. This position in the $y$ dimension represents the typical location of the electric field minimum above the electrode surface in the initial straight section of the beam splitter. The minimum rises slightly, by $\sim$300~$\mu$m, in the region where the center-conductor electrodes widen before splitting. The spatial separation in the $x$ dimension of the two arms of the beam splitter at the end of the device is 13 mm.

For the Rydberg-Stark states prepared in the experiments with electric dipole moments of $\mu_{\mathrm{elec}}$ = 8700~D, the depth of the guide when $V_{\mathrm{g}}$ = 30~V is $E$ = 2$\times$10$^{-22}$~J ($E/{k_\mathrm{B}}$ $\simeq15$~K). Under these conditions the transverse trap frequency in the device is $\Delta E/h \simeq$ 90~kHz ($\Delta E/{k_\mathrm{B}} \simeq  4~\mathrm{\mu}$K). The transverse temperature of the Rydberg atoms in the supersonic beam used in the experiments described here is $\simeq$ 1~mK (see Sec.~\ref{sec:results}). Therefore many quantized motional states of the beam splitter are expected to be populated. By miniaturizing the electrode structure and increasing $V_{\mathrm{g}}$, it is expected that trap frequencies exceeding 2~MHz ($\Delta E/{k_\mathrm{B}} \simeq$ 0.1~mK) could be achieved. This would suggest that with larger fields and careful matching of the phase-space properties of the atomic beam to the beam splitter, only the lowest transverse motional states would be populated offering opportunities for Rydberg atom interferometry experiments.

\begin{figure}
\begin{center}
\includegraphics[width = 0.48\textwidth, angle = 0, clip=]{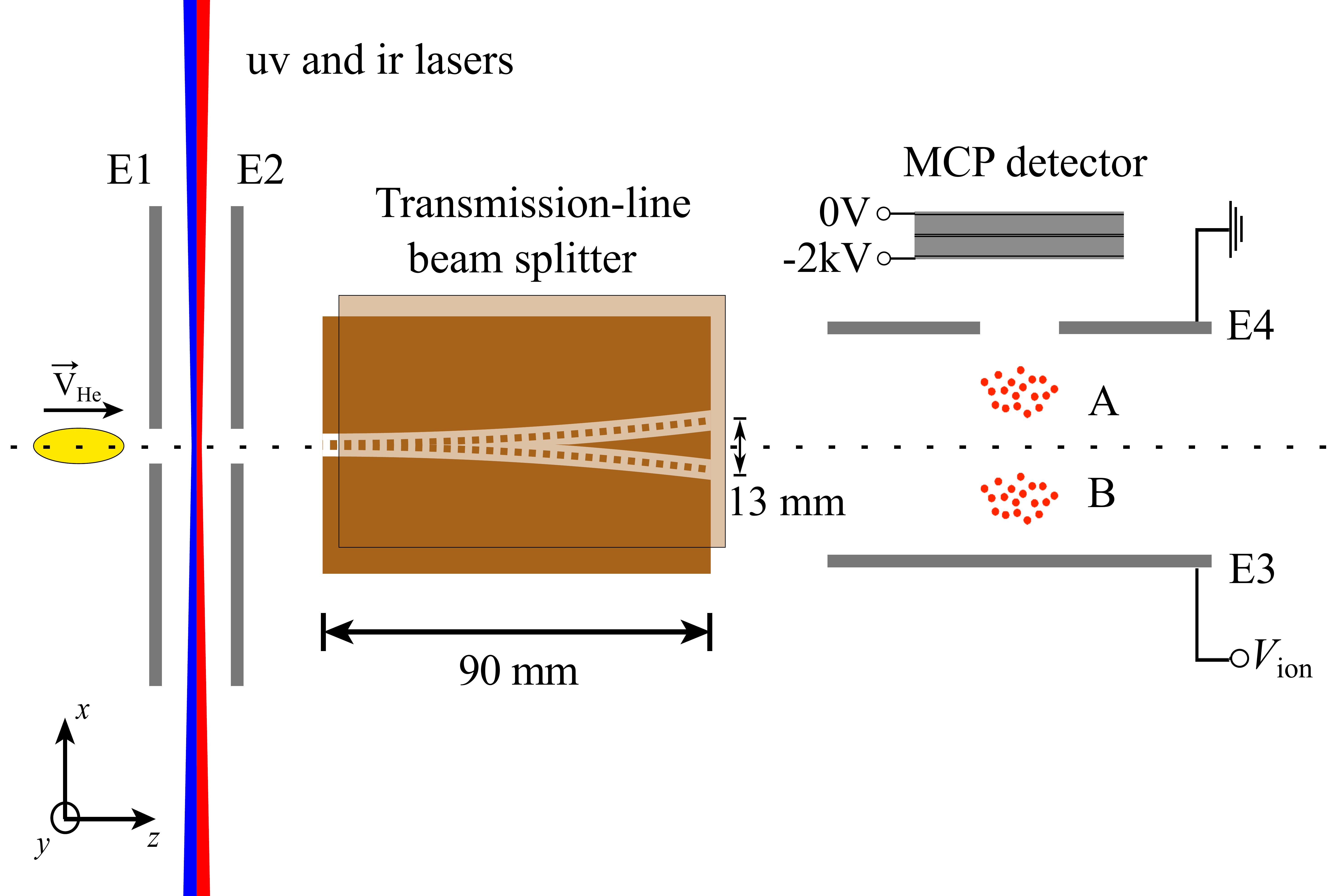}
\caption{(Color online) Schematic diagram of the experimental setup (not to scale). Rydberg state photoexcitation occurs between electrodes E1 and E2. The excited Rydberg atoms are detected by pulsed electric field ionization upon the application of a pulsed potential  $V_{\mathrm{ion}}$ = +500~V to E3 with the resulting He$^+$ ions accelerated through the aperture in E4 to the microchannel plate (MCP) detector.}
\label{fig2}
\end{center}
\end{figure}

\section{EXPERIMENT}\label{sec:experiment}

The location of the beam splitter within the experimental apparatus can be seen in Fig.~\ref{fig2}. The experiments were performed using a supersonic beam of metastable helium (He) atoms generated in a dc electric discharge at the exit of a cooled, pulsed valve. The typical mean longitudinal velocity of the beam was $\overline{v}_{\mathrm{He}}=$~1700~m/s. After passing through a skimmer, the atoms were photoexcited to Rydberg states with $n$=52 using the resonant 1s2s\,$^3\mathrm{S}_1$ $\rightarrow$ 1s3p\,$^3\mathrm{P}_2$ $\rightarrow$ 1s$n$s/1s$n$d two-color two-photon excitation scheme using focused, copropagating cw laser beams in the ultraviolet ($\lambda_\mathrm{uv}$ $\simeq$ 388.9751~nm) and infrared ($\lambda_\mathrm{uv}$ $\simeq$ 787.06~nm) regions of the electromagnetic spectrum for each step, respectively. The excitation occurred in the presence of a homogeneous electric field of 0.47~V/cm generated by applying potentials of +500~mV and -40~mV to E1 and E2 in Fig.~\ref{fig2}, respectively. This allowed the preparation of selected hydrogenic Rydberg-Stark states with $|m_{\ell}| = 2$ for which the avoided crossings at and above the Inglis-Teller limit are sufficiently small ($\Delta E/h \lesssim$ 100~kHz) to be traversed diabatically~\cite{vliegen04a,hogan16a}. The excited atoms passed through an aperture in E2 and entered the beam splitter which was operated at a potential $V_{\mathrm{g}}$. After traveling the length of the device, the atoms were detected by pulsed electric field ionization (PFI) upon the application of a pulsed potential $V_{\mathrm{ion}}$ =~+500 V to the metallic plate labeled E3 in Fig.~\ref{fig2}. The resulting He$^+$ ions were then accelerated towards a microchannel plate (MCP) detector through an aperture in electrode E4. By measuring the time between the application of the PFI potential and the arrival of the He$^+$ ions at the MCP, the distribution of atoms in the $x$ dimension at the time of ionization could be spatially resolved.

\begin{figure*}
\begin{center}
\includegraphics[width = 0.80\textwidth, angle = 0, clip=]{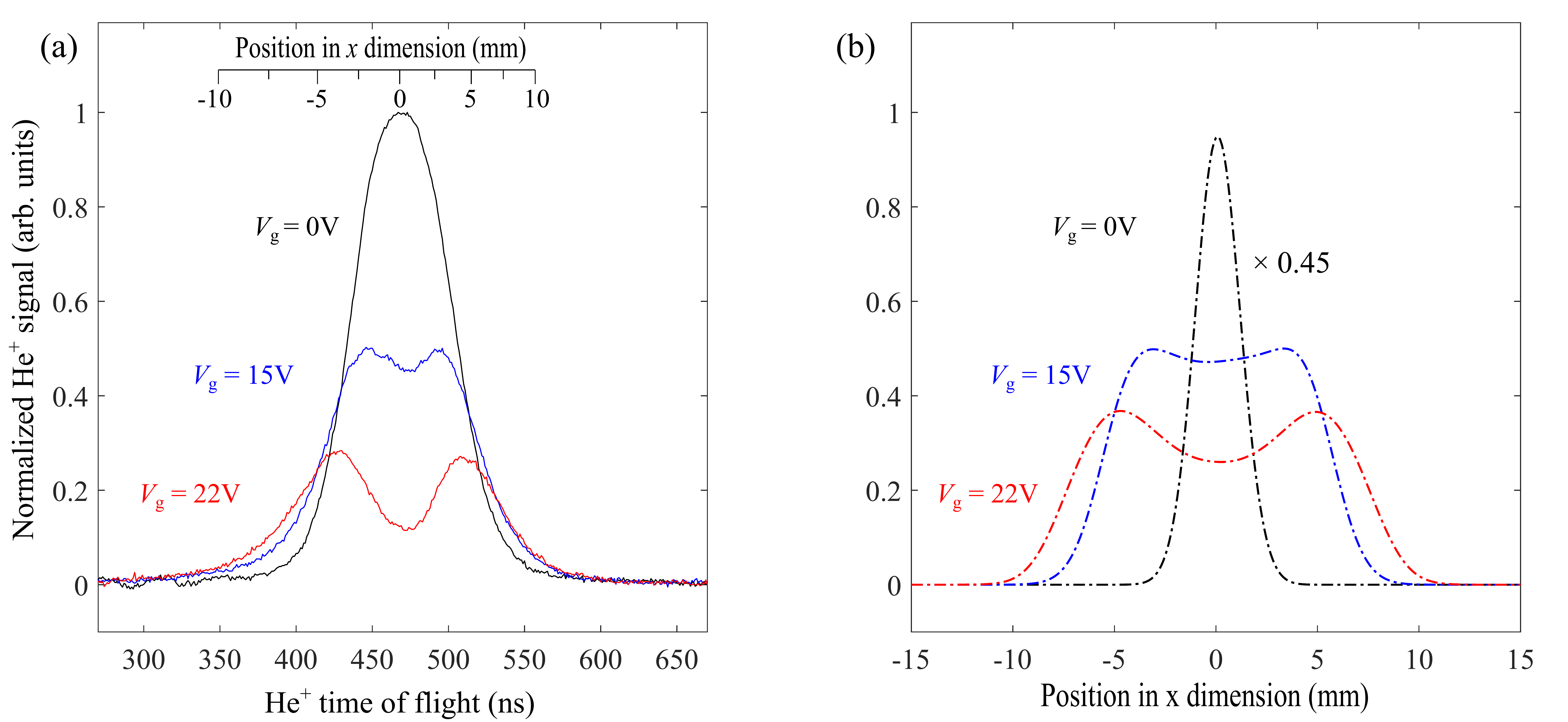}
\caption{(Color online) (a) He$^+$ time-of-flight distributions recorded following pulsed electric field ionization in the detection region of the apparatus when $V_{\mathrm{g}}$ = 0~V (black curve), 15~V (blue curve) and 22~V (red curve) as indicated. The inset axis shows the calculated relative positions of the atoms in the $x$ dimension at the time of PFI. (b) Simulated spatial distributions of Rydberg atoms in the $x$ dimension in the detection region at the time of PFI. The data for $V_{\mathrm{g}}$ = 0~V (dashed black curve) have been scaled by a factor of 0.45 as indicated. }
\label{fig3}
\end{center}
\end{figure*}

\section{RESULTS}\label{sec:results}

When the beam splitter was off, i.e., when $V_{\mathrm{g}}$ = 0~V, the trajectories of the atoms were unaffected as they traveled to the detection region along the axis of the apparatus (dashed horizontal line in Fig.~\ref{fig2}). After ionization the He$^+$ ions arrived at the MCP with a mean time of flight of $\sim$475~ns. The corresponding He$^+$ time-of-flight distribution is displayed in Fig.~\ref{fig3}(a) (black curve). When the beam splitter was activated the inhomogeneous electric field distribution split the beam of atoms into two separated components labeled A and B in Fig.~\ref{fig2}. Upon ionization the He$^+$ ions associated with component A had a reduced time of flight to the MCP while those associated with component B had an increased time of flight compared to the undisturbed beam. For the Rydberg-Stark states prepared in the experiments these two spatially separated components of the beam caused a broadening of the He$^+$ time-of-flight distribution for potentials of $V_{\mathrm{g}} \gtrsim$ 10~V. This can also be seen in the data in Fig.~\ref{fig3}(a). When operated at $V_{\mathrm{g}}$ = 22~V almost complete separation of the two split components of the beam is seen with peaks in the He$^+$ time-of-flight distribution at $\sim$425~ns and $\sim$525~ns.

The spatial separation of the split components of the Rydberg atom beam in Fig.~\ref{fig3}(a) was determined by comparing the relative flight times of the corresponding He$^+$ ions to the MCP detector, with numerical calculations of the dependence of the He$^+$ time of flight on the position of ionization of the He Rydberg atoms. This comparison resulted in the position scale included in the top part of Fig.~\ref{fig3}(a) which indicates a separation of 8.5~mm between the two split components of the beam when $V_{\mathrm{g}}$ = 22~V.

To obtain further insight into the Rydberg atom trajectories from their time of photoexcitation to the time of electric field ionization at the end of the beam splitter numerical simulations of particle trajectories through the complete apparatus were performed. In these simulations the spatial dimensions of the initially excited Rydberg atom ensemble were set by the parameters of the photoexcitation laser beams to be 0.4 and 0.1~mm full-width-at-half-maximum in the $x$ and $y$ dimensions, respectively, while the length of the excited Rydberg atom cloud was 6~mm in the $z$ dimension. The corresponding relative translational temperatures of the atoms were 1.7, 0.7 and 78~mK in the $x$, $y$ and $z$ dimensions, respectively. The results of these simulations for $V_{\mathrm{g}}$ = 0, 15 and 22~V are displayed in Fig.~\ref{fig3}(b). For ease of comparison with the data in Fig.~\ref{fig3}(a) the calculated spatial distribution for $V_{\mathrm{g}}$ = 0~V in Fig.~\ref{fig3}(b) has been scaled by 0.45. The calculated spatial distributions of Rydberg atoms at the detection region of the apparatus show a similar dependence on $V_{\mathrm{g}}$ to the experimental data with a clear separation between the two split components of the beam when $V_{\mathrm{g}} \gtrsim$ 15~V. From the results of these simulations, the reduction in signal with increasing values of $V_{\mathrm{g}}$ observed in the experimental data in Fig.~.3(a) was identified to originate from vertical deflection of the beam of atoms into the $y$ dimension. The differences in the relative amplitudes of the experimentally measured He$^+$ time-of-flight distributions and calculated data are attributed to the effects of state-changing collisions with background gas from the beam-splitter substrate~\cite{lancuba16a}. These collisions do not strongly affect the signal amplitude when the beam splitter is inactive but contribute to the loss of atoms when $V_{\mathrm{g}} >$ 0~V. The difference in the relative widths of the experimental and calculated data, particularly when $V_{\mathrm{g}}$ = 0~V, is the result of the nonlinearity of the MCP response.

\begin{figure}
\begin{center}
\includegraphics[width = 0.48\textwidth, angle = 0, clip=]{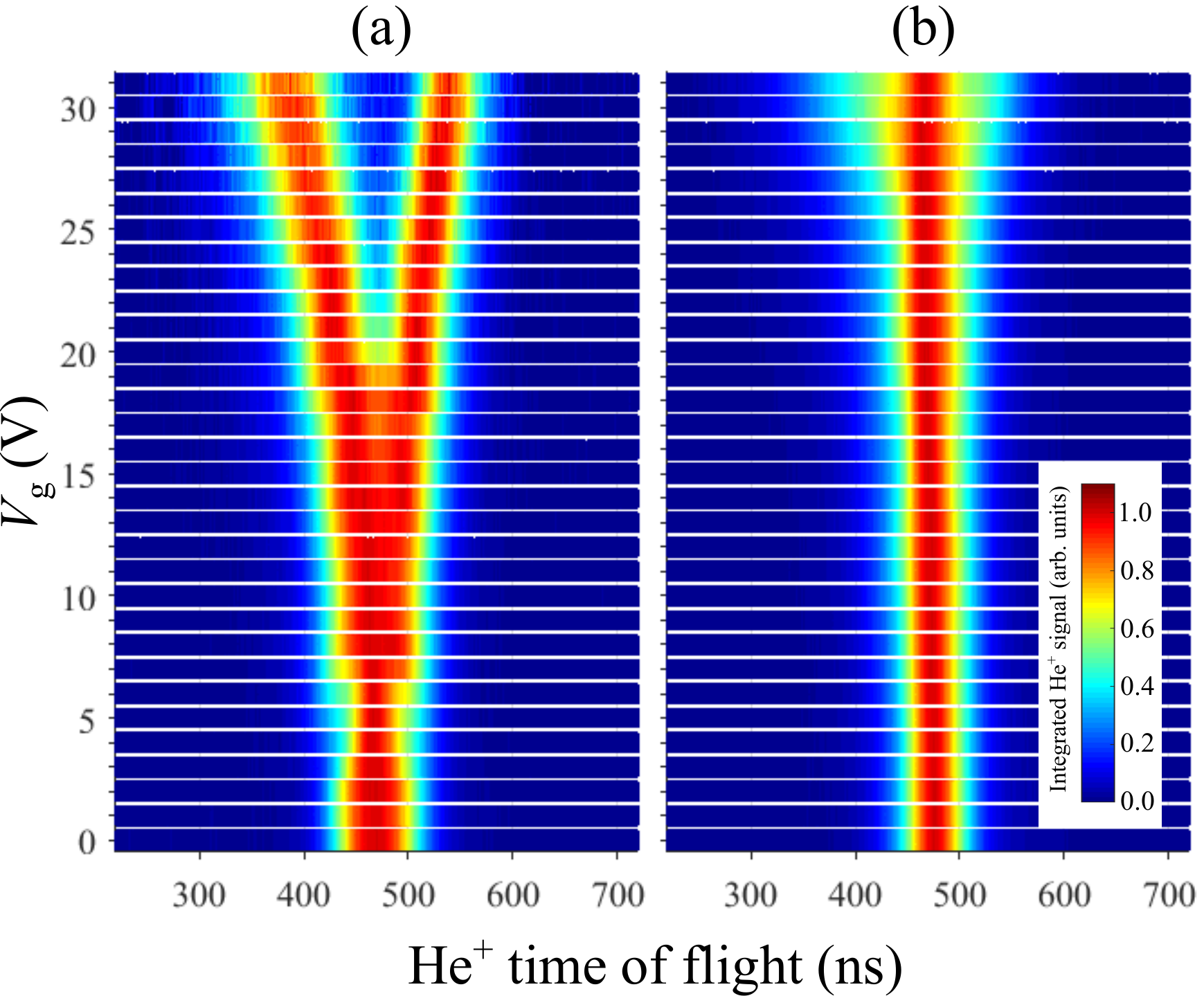}
\caption{(Color online) Normalized He$^+$ time-of-flight distributions for a range of beam-splitter operating potentials, $V_{\mathrm{g}}$, for (a) a low-field-seeking Rydberg-Stark state for which $\mu_{\mathrm{elec}}$ = 8700~D, and (b) a center Stark state with $\mu_{\mathrm{elec}}\simeq0$~D. The color scale in (b) is common to both plots.}
\label{fig4}
\end{center}
\end{figure}

To fully characterize the operation of the beam splitter further measurements were made for a range of operating potentials, $V_{\mathrm{g}}$. The results of these measurements are shown in Fig.~\ref{fig4} (a) with the intensity maximum in each data set normalized to unity. For low potentials, i.e., when 0~V $\leq V_{\mathrm{g}} \leq$ 7~V, the He$^+$ time-of-flight distributions are seen to gradually narrow. This indicates that under these conditions the initial straight section of the guide transversely focuses the atoms to the detection region. For higher potentials, i.e., $V_{\mathrm{g}} >$ 8~V, the time-of-flight distributions broaden and begin to split into two components. A clear spatially resolved splitting is seen when $V_{\mathrm{g}} >$  20 V. For the largest potential applied, $V_{\mathrm{g}}$ = 31~V, the splitting in the time-of-flight distribution corresponds to a spatial separation of the two components of the beam in the $x$ direction at the time of PFI of 15.6~mm. The slight asymmetry of the split components of the beam arises because the ion flight times are not directly proportional to the position of the atoms in the $x$ dimension at the time of PFI (see axis inset in Fig.~\ref{fig3}).

To demonstrate that the beam splitter only operates to split beams of atoms in low-field-seeking Rydberg-Stark states the measurements were repeated following photoexcitation of states in the center of the Stark manifold with $n$=52 and electric dipole moments of $\mu_{\mathrm{elec}}\simeq$ 0~D. The resulting data are displayed in Fig.~\ref{fig4}(b). It can be seen from these measurements that the atoms in these states are generally unaffected by the electric fields of the beam splitter. However, when operated at high potentials, e.g., $V_{\mathrm{g}} >$ 20 V, a small amount of broadening in the He$^+$ time-of-flight distributions is observed. This broadening is attributed to deflection of atoms pumped into Stark states with small nonzero electric dipole moments by blackbody radiation during their $\sim$45~$\mu$s flight through the device.

\section{ELECTRIC FIELD MODULATION}\label{sec:electricfield}

\begin{figure}
\begin{center}
\includegraphics[width = 0.48\textwidth, angle = 0, clip=]{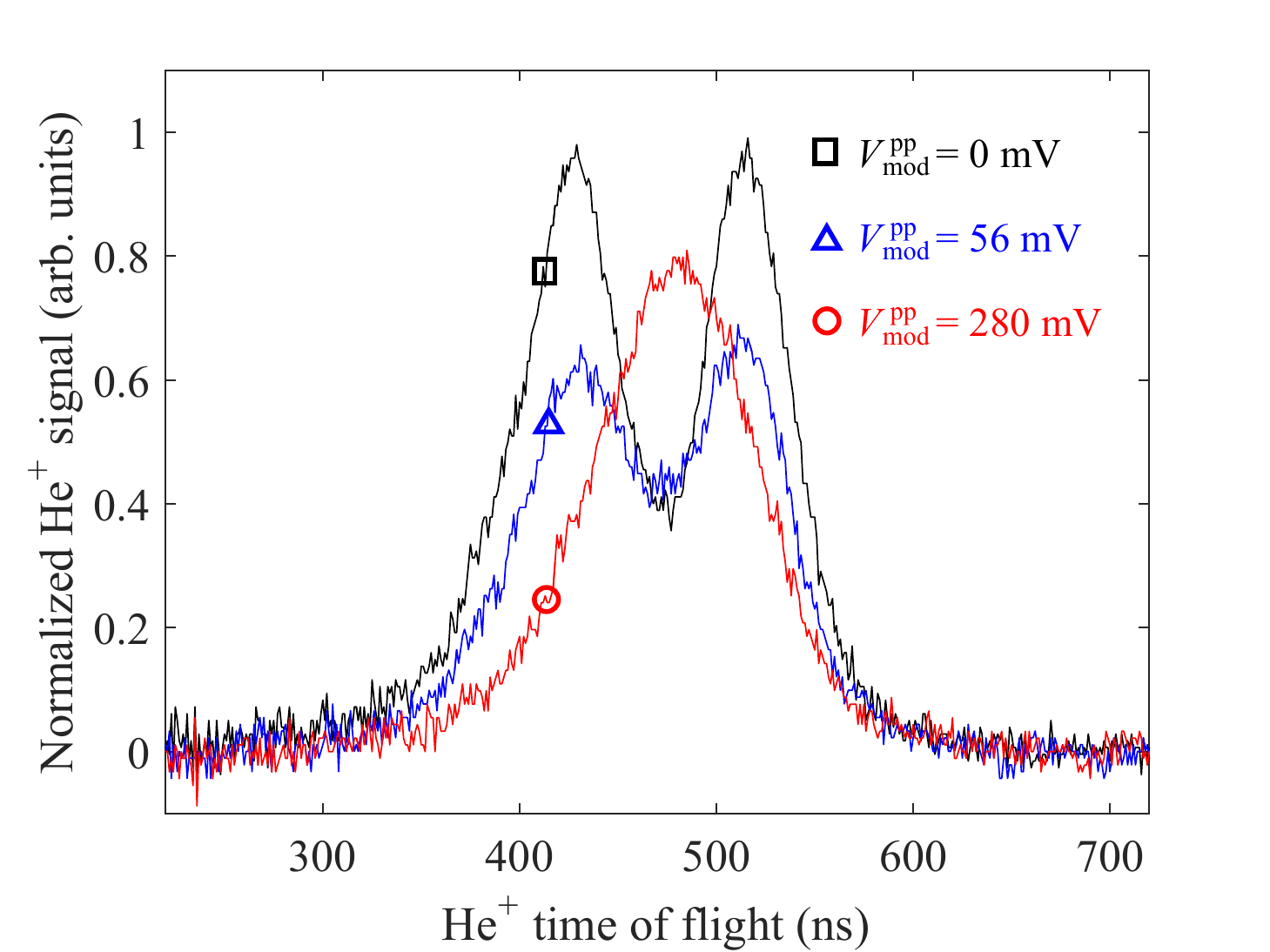}
\caption{(Color online) He$^+$ time of flight distributions for $V_{\mathrm{g}}$ = 20~V with no amplitude modulation (black curve) and with modulations of $V_{\mathrm{mod}}^{\mathrm{pp}}$ = 56 and 280~mV (blue and red curves, respectively) at 100~MHz .}
\label{fig5}
\end{center}
\end{figure}

The sensitivity of the beam splitter to changes in the electric dipole moments of the atoms during their trajectories through the device, as seen in Fig.~\ref{fig4}(b), can be further exploited to identify processes that contribute to state changing in this and other Rydberg atom optics elements. In particular this sensitivity can be used to investigate the effects of modulation of the electric fields of the beam splitter by periodic driving, or laboratory noise, on the evolution of the trapped or guided atoms. To observe these effects the atoms were prepared in the outer low-field-seeking Rydberg-Stark state and the device was operated at a potential of $V_{\mathrm{g}}$ = 20~V. A small amplitude modulation was applied to this potential at a frequency of 100~MHz. The effect of this modulation with peak-to-peak amplitudes in the range from $V_{\mathrm{mod}}^{\mathrm{pp}}$ = 0 - 280~mV on the split He$^+$ time-of-flight distributions can be seen in Fig.~\ref{fig5}. For these measurements $\overline{v}_{\mathrm{He}}$ = 2000~m/s and the center black set of data was recorded with $V_{\mathrm{mod}}^{\mathrm{pp}}$ = 0~V as indicated. When $V_{\mathrm{mod}}^{\mathrm{pp}}$ is increased to 56~mV (blue data set), the amplitude of the signal from the split components of the beam reduces. This observation suggests that the modulation of the depth of the guide leads to heating of the atoms or state-changing into Stark states with lower electric dipole moments that are less affected by the inhomogeneous electric field. This effect becomes more exaggerated when $V_{\mathrm{mod}}^{\mathrm{pp}}$ is increased to 280~mV. Introduction of similar modulations to the electric potentials in the particle trajectory simulations, to account for their effect on the Rydberg atom dynamics but not state-changing, indicated no change in the efficiency of the beam splitter. This suggests that the dominant loss mechanism associated with modulation of the beam-splitter potentials results from state-changing. In the absence of this imposed amplitude modulation, electrical laboratory noise~\cite{zhelyazkova15a,zhelyazkova15b} can also have a similar detrimental effect on the efficiency with which this and other Rydberg atom optics elements are operated. Such noise must also be considered in accounting for contributions to $m$-changing which is of importance on longer time scales in these experiments ~\cite{seiler16a}.

\section{CONCLUSION AND OUTLOOK}\label{sec:conclusion}

We have demonstrated the operation of an electrostatic beam splitter for atoms in hydrogenic Rydberg-Stark states with large static electric dipole moments.  The two-dimensional array of electrodes used in the design of this device is scalable, and permits the integration of the beam splitter with transmission-line guides, decelerators and traps for Rydberg atoms, and with co-planar microwave waveguides and resonators. 

In the experiments reported here supersonic beams of He atoms were split into pairs of separated components of equal intensity. By adjusting the electrical potentials, or the electrode dimensions in each arm of the beam splitter it is foreseeable to construct devices in which the intensity ratios of the split components of the beams can be modified. This will be a valuable control parameter, for example, when using one of the split components of a beam as an intensity reference in studies of resonant energy transfer in collisions of Rydberg atoms with polar ground state molecules~\cite{zhelyazkova17a}. 

For the present configuration of electrodes and an operating potential of $V_{\mathrm{g}}=30$~V, beams of He atoms for which $\overline{v}_{\mathrm{He}}=1700$~m/s can be split using this beam splitter when prepared in Stark states with electric dipole moments exceeding 3850~D. Therefore the device can be employed for a large number of Rydberg states with values of $n>33$.

The observed loss of guided atoms upon amplitude modulation of the beam splitter operating potentials highlights the effects that undesirable time-dependent fields, such as those associated with electrical laboratory noise, can have on the efficient operation of Rydberg atom optics elements. It will be important to consider effects of this kind in the development of more complex arrays of tools for controlling the translational motion of cold Rydberg atoms and molecules, and in the interpretation of measurements of the decay of trapped atoms and molecules. 

Finally, through a combination of collimation (or transverse cooling) of the atomic He beam, and increasing the electric field gradients in the beam splitter by operating it at higher electrical potentials or scaling down the sizes of the electrode structures, we expect that it will become possible to inject atoms into a small number of the lowest-lying, quantized, transverse motional states of the device. This would allow intriguing new opportunities for Rydberg atom interferometry which could also be exploited in experiments with positronium atoms. 

\begin{acknowledgments}
This work was supported financially by the Department of Physics and Astronomy and the Faculty of Mathematical and Physical Sciences at University College London, and the Engineering and Physical Sciences Research Council under Grant No.~EP/L019620/1.
\end{acknowledgments}

\end{document}